# Fluctuation of TeV to EeV Energy Muons and Induced Muon Showers in Water

Y. Okumura, N. Takahashi
*Graduate School of Science and Technology, Hirosaki University, Hirosaki 036-8561, Japan*
A. Misaki
*Innovative Research Organization, Saitama University, Saitama 338-8570, Japan*

By using the integral method in the muon propagation through water, we calculate the range fluctuation of high and ultra high energy muons. Many authors divide all radiative processes into two parts, namely, the continuous part and radiative part in their Monte Carlo simulation in order to consider the fluctuation in the both ranges and energies of the muons, while we treat all stochastic processes as exactly as possible, without the introduction of the continuous parts in all stochastic processes. The validity of our Monte Carlo method is checked by the corresponding analytical method which is methodologically independent on the Monte Carlo procedure. Accompanied cascade showers are generated by the direct electron pair production, bremsstrahlung and photo-nuclear interaction. These showers are calculated by the exact Monte Carlo Method in one dimensional way. We report survival probabilities, range distributions and examples of individual muon behavior.

## 1. INTRODUCTION

The range fluctuation in high energy muons behavior may play an important role in the analysis of muon neutrino events for KM3 detector deployed in the Antarctic, the ocean and the lake [1-4]. As far as the treatment of the range fluctuation of high energy muons are concerned, there are two independent Monte Carlo methods. The one is the differential method in which the muons concerned are pursued in step by step way [5-10]. In this method, the introduction of $v_{cut}$ is essential for the separation of the radiative part from the continuous part in the stochastic processes. The other is the integral method in which the interaction points of the muons and the dissipated energies by muons are directly determined [11] and the stochastic processes are treated as exactly as possible without introduction of $v_{cut}$ into the stochastic processes. As far as the behaviors of high energy muons are concerned, these two methods are independent ones and are equivalent to each other, giving the same results. However, it should be noticed that the energy determination of the high energy muons are carried out through the measurement of the Cherenkov light which are produced by the accompanied cascade shower due to radiative processes of the muon concerned. Related to the Cherenkov light due to the accompanied cascade showers, some difference will appear in the results due to the differential method and due to the integral one.

In this paper, the diversity of the behaviors of the muons is shown in the light of the cascade showers which are the origin of the Cherenkov light.

## 2. THE COMPARISON OF OUR CALCULATION WITH THE OTHERS

Our results by the integral method are compared with the results by the differential method in order to confirm the validity of our method. In Figure 1, we compare our survival probabilities of muons in the standard rock for given initiated energies with the corresponding ones given by Lipari and Stanev [5]. The agreements are rather well.

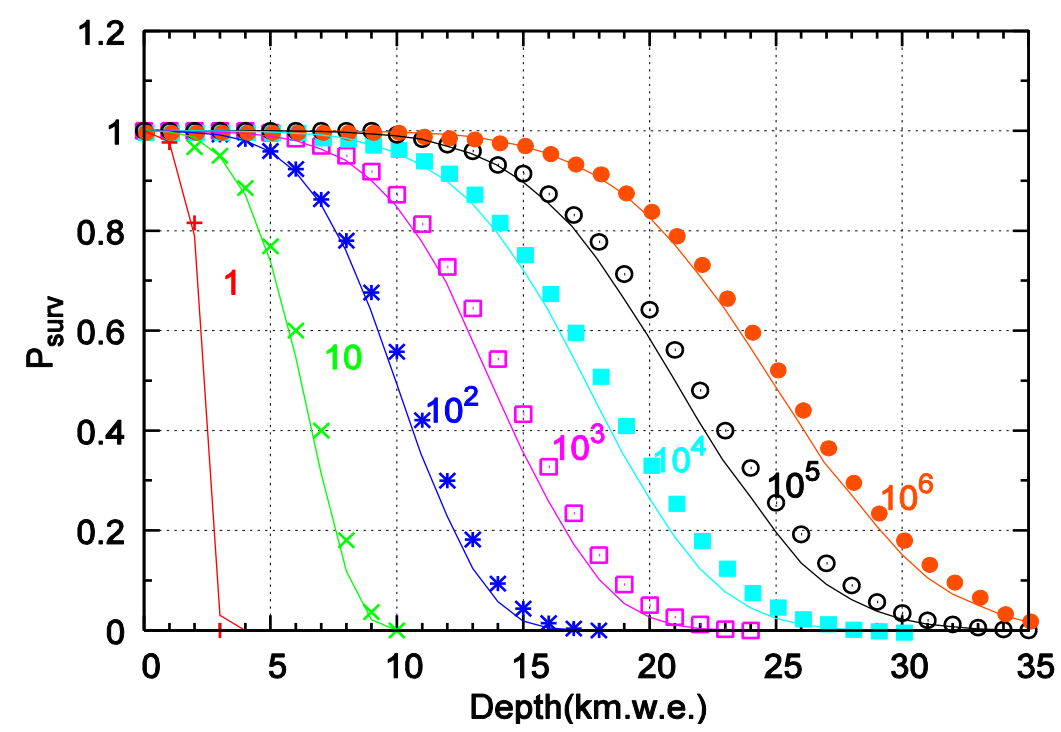

Figure 1: The comparison of our results with Lipari and Stanev in the survival probabilities as the function of the depth for 1TeV to 1EeV. Curves show ours calculations, while symbols show Lipari and Stanev's ones.

## 3. RANGE DISTRIBUTIONS FOR THE DIFFERENT INITIATED ENERGIES

Based on the validity of our solutions by the integral method which is proved to be correct by other differential method as shown in Figure 1, we give the range distributions for initiated energies of 1TeV, 1PeV and 1EeV in Figures 2, 3 and 4, respectively. The minimum energy of the muon is taken as 1GeV.

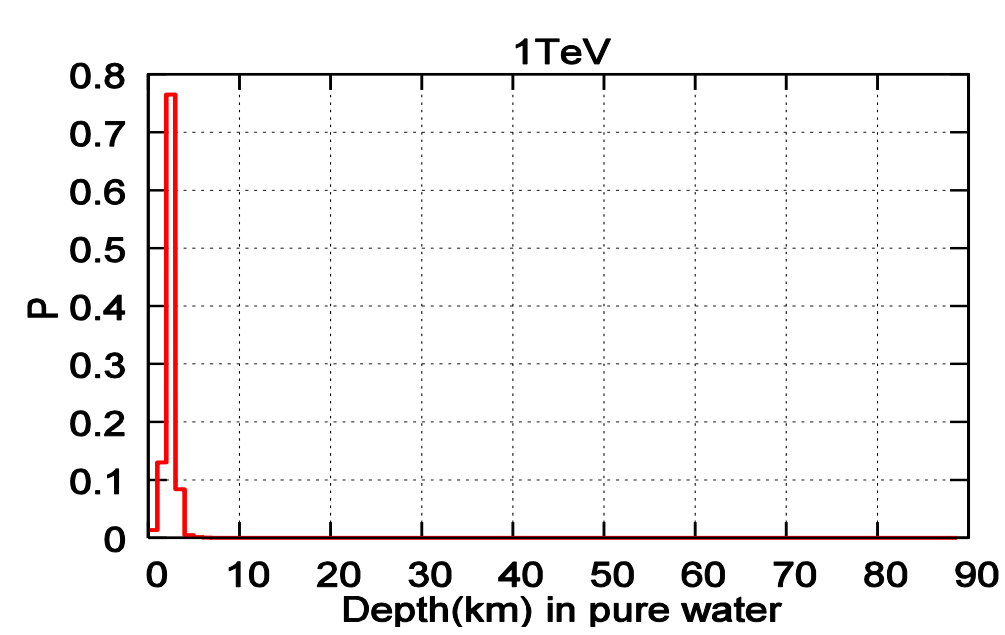

Figure 2: The range distribution for initiated energy of 1TeV.

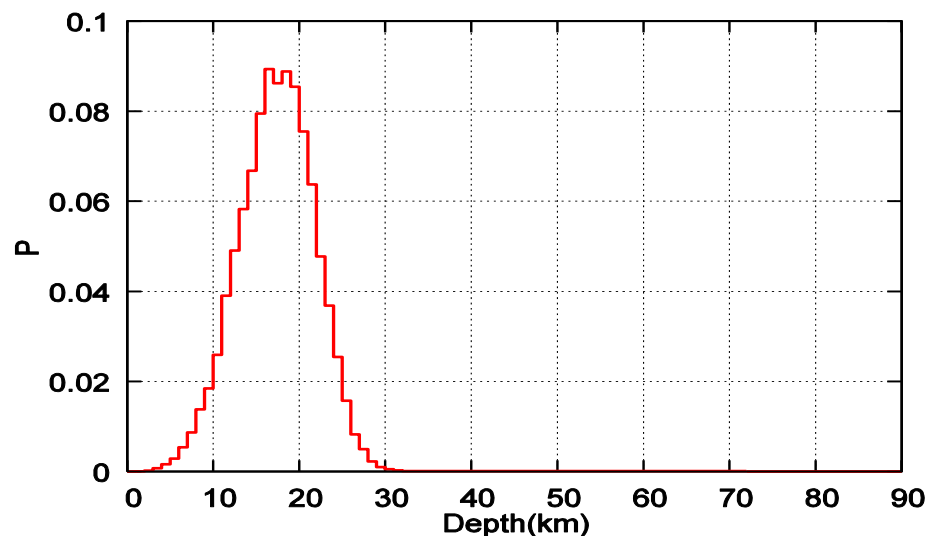

Figure 3: The range distribution for the initiated energy of 1PeV.

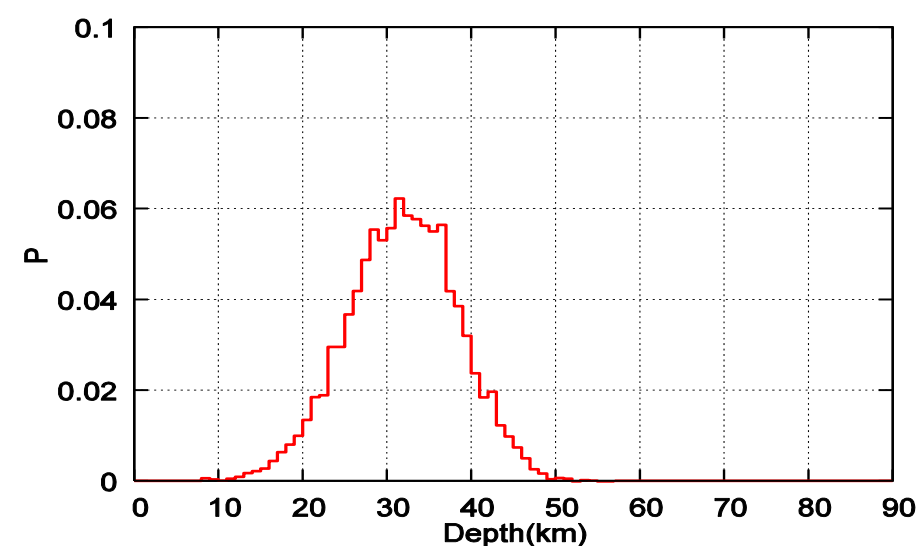

Figure 4: The range distribution for the initiated energy of 1EeV.

$P(R;E_p)$, probability for these range distributions can be quite well approximated as the normal distribution in the following.

$$P(R;E_p) = \frac{1}{\sqrt{2\pi}\sigma} \exp\left( -\frac{R - \langle R \rangle}{2\sigma^2} \right) \quad (1)$$

, where $E_p$, $R$, $\langle R \rangle$ and $\sigma$ are the initiated energy, range of muon, average value of the ranges and their standard deviation, respectively and these quantities are given in Table1 for the initiated energies of $10^{11}$ eV to $10^{18}$ eV.

In order to show the diversity in the muon's behavior with the same initiated energy, in Figures 7 to 9, we give the typical behavior of individual muon with regard to their energy losses. Namely, we show a muon with the shortest range, a muon with average-like range and a muon with the longest range among 100,000 sampled events with the same initiated energy of 1PeV.

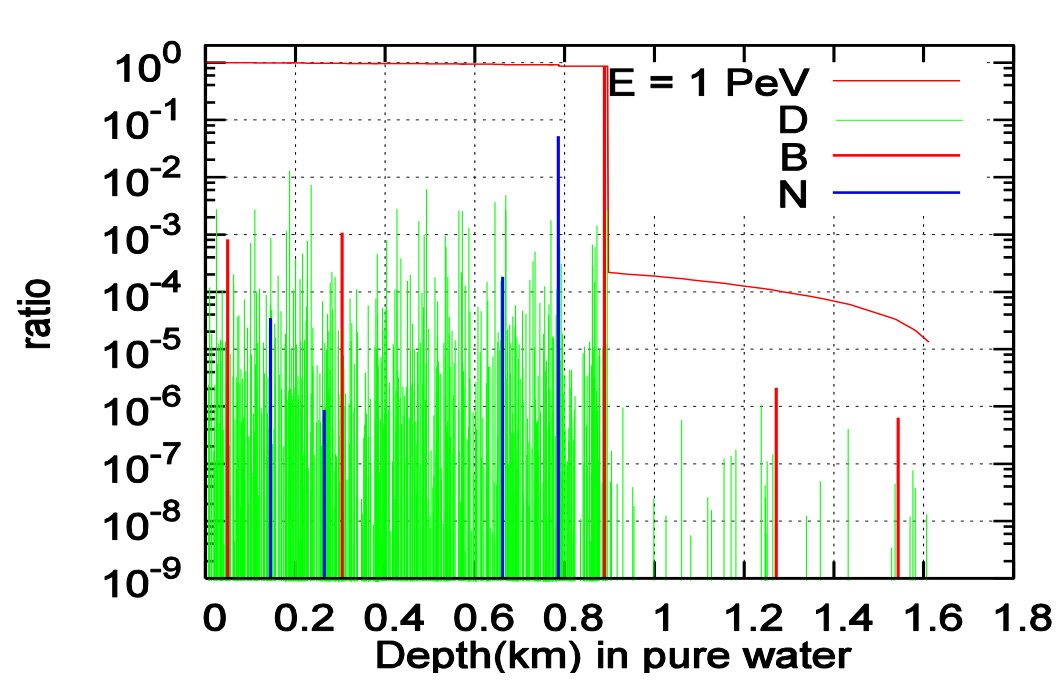

Figure 7: The example of the muon with the shortest range for 1PeV among 100,000 sampled events. The energy losses due to each radiative process in individual muon as the function of the depth traversed. The "ratio" in ordinate axis denotes the ratio of each energy loss to initiated energy. D, B and N denote each energy loss due to direct pair production, bremsstrahlung and nuclear interaction, respectively.

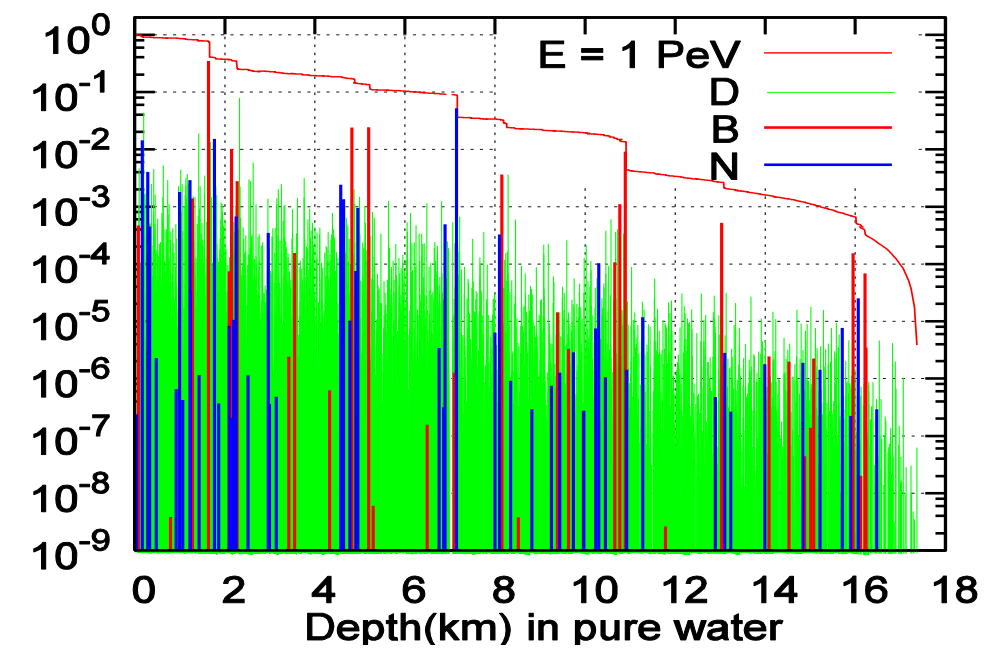

Figure 8: The example with the average like range. The other symbols are the same as in Figure 7.

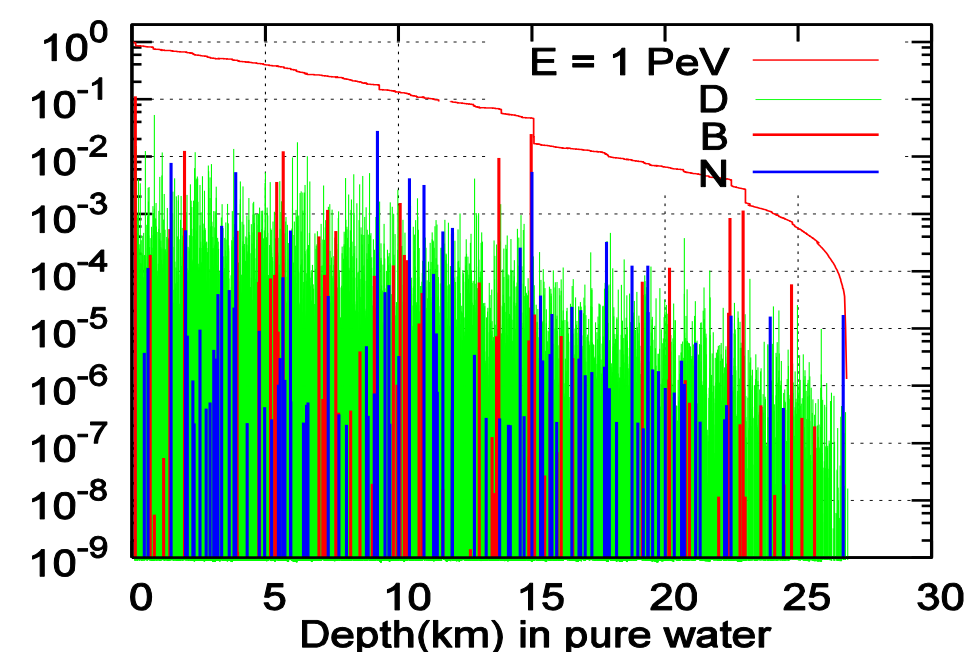

Figure 9: The example with the longest range. The other symbols are the same as in Figure 7.

In Table 2, we summarize the interrelation in the energy loss among three radiative processes for the shortest, the average-like and the longest ranges for 1TeV, 1PeV and 1EeV respectively. Numerical values under each radiative process denote the ratio of the energy loss to the total energy loss. It is easily understood from the table that the muons with the shortest range lose their energies catastrophically by bremsstrahlung, the muon with the longest lose their energy by a lot of direct pair production with smaller energy, and the muons with the average-like lose their pretty part by both direct pair production and bremsstrahlung.

## 4. THE CORRELATION BETWEEN DISIPATED ENERGY DUE TO MUONS AND RESULTANT CASCADE SHOWERS

The high energy muons lose their energy through the bremsstrahlung, direct pair production and nuclear interaction.

The bremsstrahlung process produces photon-initiated cascade shower, the direct pair production produces a pair of electron initiated cascade showers and nuclear interaction process produces the hadron showers, the

.

Table 1:

| $E_p$ $[eV]$ | $\langle R \rangle$ $[g/cm^2]$ | $\langle R \rangle$ $[km]$ | $\sigma$ $[g/cm^2]$ | $\sigma$ $[km]$ |
|---|---|---|---|---|
| $10^{11}$ | $4.75\times10^{4}$ | $4.75\times10^{-1}$ | $2.39\times10^{4}$ | $2.39\times10^{-1}$ |
| $10^{12}$ | $2.51\times10^{5}$ | $2.51\times10^{0}$ | $5.32\times10^{4}$ | $5.32\times10^{-1}$ |
| $10^{13}$ | $7.18\times10^{5}$ | $7.18\times10^{0}$ | $1.97\times10^{5}$ | $1.97\times10^{0}$ |
| $10^{14}$ | $1.23\times10^{6}$ | $1.23\times10^{1}$ | $3.35\times10^{5}$ | $3.35\times10^{0}$ |
| $10^{15}$ | $1.73\times10^{6}$ | $1.73\times10^{1}$ | $4.35\times10^{5}$ | $4.35\times10^{0}$ |
| $10^{16}$ | $2.22\times10^{6}$ | $2.22\times10^{1}$ | $5.16\times10^{5}$ | $5.16\times10^{0}$ |
| $10^{17}$ | $2.70\times10^{6}$ | $2.70\times10^{1}$ | $5.86\times10^{5}$ | $5.86\times10^{0}$ |
| $10^{18}$ | $3.19\times10^{6}$ | $3.19\times10^{1}$ | $6.52\times10^{5}$ | $6.52\times10^{0}$ |

Table 2:

| $E_p = 10^{12} eV$ | **Brems** | **Direct Pair** | **Nuclear** |
|---|---|---|---|
| <Average> | $3.37\times10^{-1}$ | $5.26\times10^{-1}$ | $1.37\times10^{-1}$ |
| Shortest | $9.99\times10^{-1}$ | $4.81\times10^{-4}$ | $4.06\times10^{-4}$ |
| Average-like | $2.94\times10^{-1}$ | $3.62\times10^{-1}$ | $3.44\times10^{-1}$ |
| Longest | $2.95\times10^{-1}$ | $6.97\times10^{-1}$ | $8.04\times10^{-3}$ |
| $E_p = 10^{15} eV$ | | | |
| <Average> | $3.40\times10^{-1}$ | $4.98\times10^{-1}$ | $1.62\times10^{-1}$ |
| Shortest | $8.59\times10^{-1}$ | $8.87\times10^{-2}$ | $5.20\times10^{-2}$ |
| Average-like | $4.25\times10^{-1}$ | $4.78\times10^{-1}$ | $9.72\times10^{-2}$ |
| Longest | $1.84\times10^{-1}$ | $7.57\times10^{-1}$ | $5.90\times10^{-2}$ |
| $E_p = 10^{18} eV$ | | | |
| <Average> | $3.24\times10^{-1}$ | $4.59\times10^{-1}$ | $2.17\times10^{-1}$ |
| Shortest | $7.04\times10^{-1}$ | $2.91\times10^{-1}$ | $4.77\times10^{-3}$ |
| Average-like | $4.22\times10^{-1}$ | $3.94\times10^{-1}$ | $1.84\times10^{-1}$ |
| Longest | $6.78\times10^{-2}$ | $6.86\times10^{-1}$ | $2.46\times10^{-1}$ |

most of which consist of the aggregate of photon-initiated cascade showers.

In order to demonstrate the essential of the correlation between the energy losses due to different elementary processes and the cascade showers which are produced by the corresponding elementary process, in Figure 10, we give one example of 1EeV muon which are given as the function of the depth traversed by the muon, the character of which is similar to Figures 7 to 9. In Figure 11, we give the energy losses due to the muon concerned are transformed into the electron numbers which are produced by the corresponding cascade shower. Namely, the electron numbers are given as the function of the traversed by the muon. Here, we take the minimum energy of cascade shower particle as 100 GeV temporarily, for the purpose of extracting the characteristics of a series of cascade showers produced by the dissipated energy from muon. All processes are treated as exactly as possible by the Monte Carlo method.

It is clear from Figure 10 that the muon concerned loses the maximum energy by bremsstrahlung catastrophically at ～120 meter from the starting point and loses the second maximum energy by direct pair production ( exceptional case) at ～860 meter. It is clearly seen from the Figure 11 that the corresponding electron numbers exist at the expected depths, a little later than ～120 meter and ～860 meter in Figure 11 due to the development of the corresponding cascade showers.

As well known, the energy determination of high energy muons are carried out in the measurement of the Cherenkov light, the origin of which are a series of electron number as shown in Figure 11. In the real measurement of the Cherenkov light, we can expect duller fluctuation in the real measurement of the Cherenkov light than in the cascade showers, because of longer their attenuation length.

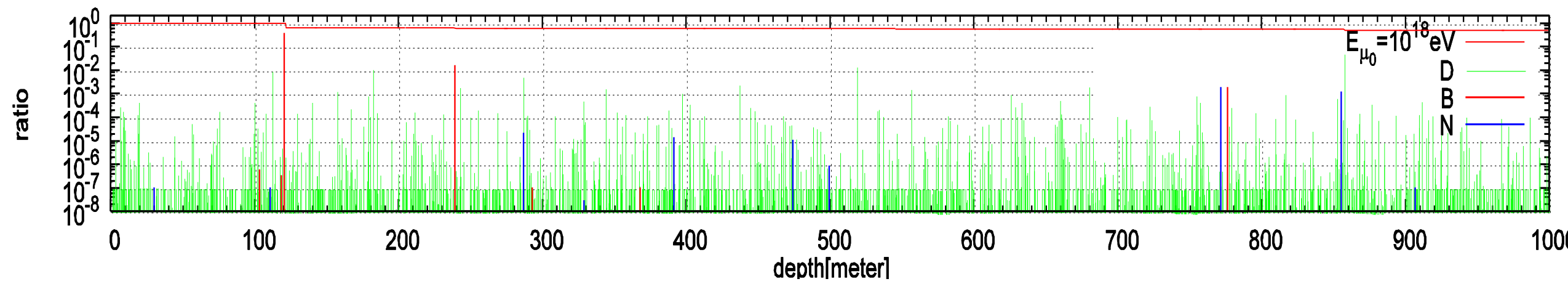

Figure 10: One example of 1EeV muon's energy losses as the function of the depth traversed. The other symbols are the same as in Figure 7.

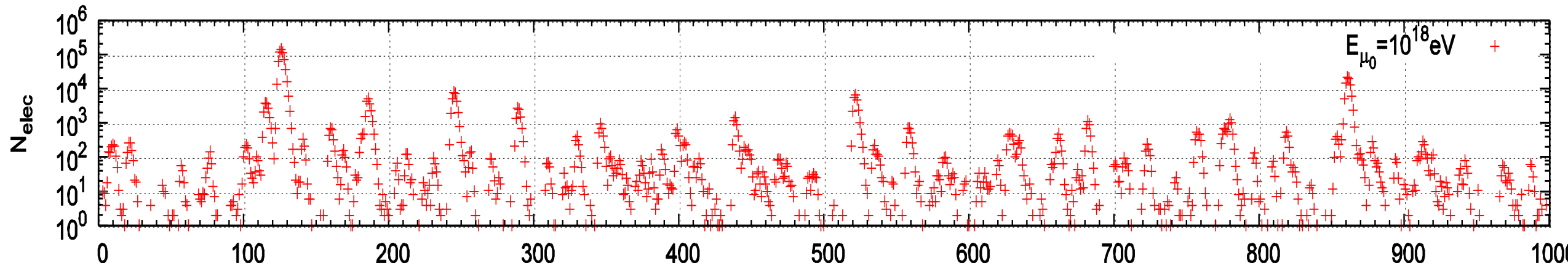

Figure 11: Electron numbers as the function of the depth traversed by the muon shown in Figure 10.

## 5. SUMMARY

The range fluctuation of the cosmic ray muons becomes strong as their energies increase. This fluctuation is directly connected with the dissipated energies due to the muons which are the origins of the Cherenkov light. Namely, the generation of the Cherenkov light due to the muon is much affected by the range fluctuation, which makes it difficult to estimate the energies of the muons in higher energies.

## References

[1] F.Halzen, S.R.Klein, Invited Review, *Sc.Instrument* (to be published)

[2] ANTARES Collab., *arXiv:1007.1777v1*, [*astro-ph.HE*] 11Juli2010

[3] NEMO Collab., *Astropart.Phys.*, **33** (2010) 263

[4] Baikal.Collab., *Astropart.Phys.*, **7** (1997) 263

[5] Paolo Lipari, Todor Stanev, *Phys.Rev.* **D44** (1991) 3543

[6] P.Antonioli et al, *Astropart.Phys.*, **7** (1997) 357

[6] S.I.Dutta et al., *Phys.Rev.* **D63** (2001) 0904020

[7] S.I.Klimushin et al., *Phys.Rev.* **D64** (2001) 014016

[8] S.Bottai, L.Perrone, *NIM in Physics Research A* **459** (2001) 319

[9] Dmitry Chirkin et al., *hep0407075v2* (2008)

[10] V.A.Kudryatsev, *Computer Physics Communications*, **180** (2009) 339

[11] N.Takahashi et al., *Proc.19th ICRC, Bangalore India*, **11** (1983) 443